\documentclass[aps,prl,twocolumn,showpacs,superscriptaddress]{revtex4-1}
\usepackage{amsfonts}

\usepackage{graphicx}
\usepackage{subfigure}
\usepackage{braket}
\usepackage{amsmath}
\usepackage{bm}

\begin{document}

\title{Efficient wavefunction propagation by minimizing accumulated action}
\author{Zachary~B.~Walters}
\affiliation{Max Born Institute, Bereich B, Max Born Strasse 2a, 12489
  Berlin Adlershof, Berlin Germany}
\date{\today}

\begin{abstract}
This paper presents a new technique to calculate the evolution of a
quantum wavefunction in a chosen spatial basis by minimizing the
accumulated action.  Introduction of a finite temporal basis reduces
the problem to a set of linear equations, while an appropriate choice
of temporal basis set offers improved convergence relative to methods based
on matrix exponentiation for a class of physically
relevant problems.
\end{abstract}

\maketitle

Calculating the time evolution of a quantum wavefunction is a
longstanding problem in computational physics.  The fundamental nature
of the problem makes it resistant to simplification.  Problems of
physical interest, such as the interaction of a molecule with a strong
laser field, may lack symmetry or involve time dependent,
nonperturbative fields.  Problems involving multiple dimensions or
multiple interacting particles may quickly grow so large as to be
unmanageable with all but the largest computational resources
\cite{taylor,feist,sansone}, a problem which is not easily outstripped
by increases in computational power.  An ideal propagator, then, must
serve two masters -- it must treat the physical side of the problem accurately,
and the computational side of the problem efficiently.


The most common approach to the problem does not treat the evolving
wavefunction directly.  Instead, the wavefunction $\psi(x,t)$ is
expanded in some chosen basis set  $\phi(x,t)=\sum_{i}c_{i}(t)
\chi_{i}(x)$, where $c_{i}(t_{0})=\braket{\chi_{i}|\psi(t_{0})}$.
Here $\psi(x,t)$ is the true wavefunction and $\phi(x,t)$ is its
representation in the chosen basis.  After this expansion has been
made, the propagation scheme may operate only on the wavefunction's
representation, rather than the wavefunction itself.
Within this general
framework, there has been a great profusion of methods for calculating
the evolution of the coefficients $c_{i}(t)$.  Popular methods include
Crank-Nicholson\cite{cranknicholson}, second order
differencing\cite{askar}, split operator\cite{bandrauk}, short
iterative Lanczos\cite{lanczos}, and Chebyshev
propagation\cite{talezer_cheby}, as well as many others\cite{dubiousways}.

As may be inferred from the large number of competing methods, the
practical question of which method works best is very difficult to
answer, and usually requires problem-specific information.
Ironically, the time dependent Schr\"odinger equation (TDSE), while
challenging to solve, is not difficult to satisfy at a particular time.
All of the above methods satisfy the TDSE exactly or approximately at
the initial time in the propagation interval.  Indeed, entire families
of propagators may satisfy the TDSE at the initial point: for the
class of propagators $\psi(x,t)(1-iH\alpha
\Delta t)=\psi(x,t+\Delta t)(1+iH(1-\alpha)\Delta t)$, $0\le \Delta
t$, the TDSE at time $t$ is satisfied to first order for any value of $\alpha$.
$\alpha=1$ yields the forwards Euler method, $\alpha=0$ the backwards
Euler, $\alpha=1/2$ the Crank-Nicholson.  Although such methods may
show sharp differences in suitability to particular problems, the TDSE
alone gives little guidance.  A full diagonalization of the
Hamiltonian matrix would satisfy the TDSE at all times; however, such
a diagonalization would be prohibitively expensive for a large
problem, and would not exist for a problem with a time dependent
Hamiltonian.   Comparison of different propagators has often involved
numerical testing on simple problems\cite{lanczoscomparison,truong} or
algorithmic scaling arguments\cite{kosloff}.

This paper addresses the problem of wavefunction propagation from the
physical perspective of minimizing the action accumulated over the chosen time
interval.  Minimizing this action is shown to be equivalent to
minimizing the time integrated error of propagation.  Because the
action is calculated over the entire time step rather than at a single
point, the constraint that it be minimized is more strict than
the TDSE, allowing the construction of a unique, variationally
optimum propagator for a particular order in time.

\section{Errors of Propagation and Representation}
A central difficulty of any numerical propagation scheme is that,
although the propagator seeks to model the evolution of an ideal
wavefunction $\psi(x,t)$, it has access only to the representation of
the wavefunction in some chosen basis, $\phi(x,t)$.  The error of the
representation is given by $\delta(x,t)=\psi(x,t)-\phi(x,t)$.

As an alternative to direct exponentiation of the Hamiltonian, a propagator
may be constructed by minimizing the integral of the error
over the time step
\begin{equation}
\text{Global error}=\int_{t}^{t+\Delta t} dt'\braket{\delta(x,t')|\delta(x,t')}.
\end{equation}
Writing the error as a two term Taylor series,
\begin{equation}
\begin{split}
\label{eq:integralequation}
\text{Global error}=\int_{t}^{t+\Delta t}dt'
\braket{\delta(x,t)|\delta(x,t)} + \\
\int_{t}^{t+\Delta t} dt'' \int_{t}^{t'}\frac{d}{dt'}
\braket{\delta(x,t')|\delta(x,t')},
\end{split}
\end{equation}
and recalling that
$i\frac{d}{dt}\psi(x,t)=H\psi(x,t)$ for the true wavefunction, the
second term in equation \ref{eq:integralequation} can be written as
\begin{equation}
\begin{split}
\label{eq:ddeltadt}
\frac{d}{dt} \braket{\delta(x,t)|\delta(x,t)}= 2 i \bra{\phi(x,t)}(i
\frac{d}{dt}-H)\ket{\phi(x,t)}+\\
 2 i \bra{\delta(x,t)}H\ket{\delta(x,t)}.
\end{split}
\end{equation}

From equations \ref{eq:integralequation} and \ref{eq:ddeltadt}, it is
apparent that the global error may arise either from imperfectly
representing the wavefunction in a particular basis (terms containing
$\delta(x,t)$) or from 
imperfectly describing the evolution of the wavefunction in that
basis (terms containing $\phi(x,t)$).  The representation error may be
minimized by an appropriate 
choice of basis; here the focus is on minimizing the error of
propagation.

The quantity $\bra{\phi(x,t)}(i\frac{d}{dt}-H)\ket{\phi(x,t)}$ found in
equation \ref{eq:ddeltadt} is the Lagrangian density, and its integral
over time gives the action accumulated by the wavefunction in a
particular interval.  However, unlike the true Lagrangian density,
here the action is calculated with respect to the representation of
the wavefunction, rather than the wavefunction itself.  The
distinction is significant.  For the true wavefunction,
minimizing the action is equivalent to setting $i \frac{d}{dt}
\psi(x,t)-H\psi(x,t)=0$ for all $x$ and $t$.  For the action
minimizing representation of the wavefunction, $|\bra{\phi(x,t)}(i
d/dt-H)\ket{\phi(x,t)}|$ is dependent on the choice of spatial and
temporal basis functions and is not guaranteed to be zero.

For a finite basis of spatial $\chi_{i}(x)$ and temporal $T_{n}(t)$
basis functions, a time dependent representation of the wavefunction
can be written as 
\begin{equation}
\phi(x,t)=\sum_{i,n}C_{in} \chi_{i}(x) T_{n}(t).
\end{equation}
In this basis, the global error of Eq. \ref{eq:ddeltadt} becomes a
function of the coefficients $C_{in}$ and the matrix representations
of the quantum mechanical operators.  Writing the Hamiltonian as the
sum of time independent and time dependent operators
\begin{equation}
H=H_{0}(x)+V(x,t)
\end{equation}
and defining the matrices
\begin{equation}
H_{ij}=\int dx \chi_{i}^{*}(x) H_{0} \chi_{j}(x)
\end{equation}
\begin{equation}
V_{ijnm}=\int_{t}^{t+\Delta t}dt' \int dx \chi_{i}^{*}(x) T^{*}_{n}(t)
V(x,t) \chi_{j}(x) T_{m}(t)
\end{equation}
\begin{equation}
O_{ij}=\int dx \chi_{i}^{*}(x)\chi_{j}(x)
\end{equation}
\begin{equation}
U_{nm}=\int dt T^{*}_{n}(t)T_{m}(t)
\end{equation}
\begin{equation}
Q_{nm}=\int dt T^{*'}_{n}(t)T_{m}(t),
\end{equation}
the change in action resulting from $C_{in}\rightarrow
C_{in}+\epsilon_{in}$ is given by
\begin{equation}
\delta S = \sum_{i,j,n,m} C_{in}[i
  O_{ij}Q_{nm}-H_{ij}U_{nm}-V_{ijnm}]\epsilon^{*}_{jm} 
\end{equation}
and the condition to minimize the accumulated action is that either
$\epsilon^{*}_{jm} =0$ (for the initial conditions) or 
\begin{equation}
\delta S_{jm} = \sum_{i,n} C_{in}[i
  O_{ij}Q_{nm}-H_{ij}U_{nm}-V_{ijnm}]\epsilon^{*}_{jm}=0
\label{eq:stationaryphase}
\end{equation}
for all j,m.  In these equations, $i$ appearing as a subscript is
treated as an index, while $i$ multiplying $O_{ij}Q_{nm}$ is the
square root of negative one.

Equation \ref{eq:stationaryphase} is the main result of this paper.
In order to construct a least action propagator, it is necessary only
to choose an appropriate temporal basis.  While in principle this
analysis applies equally well to any choice of basis, an obvious
choice is for $T_{n}(t)$ to be a set of linearly independent low-order
polynomials in t.

Lagrange interpolating polynomials provide a convenient set of
temporal basis functions.  For an evenly spaced grid $t_{m}=t+\Delta t
*m/n$ for $m=0,n$, the interpolating 
polynomials are given by 
\begin{equation}
T_{m}(t)=\Pi_{k=0,n; k \ne m} \frac{t-t_{k}}
{t_{m}-t_{k}}.
\end{equation}  
This yields a basis set of $n$ linearly independent
$n$-order polynomials in t, with the property that
$\phi(x,t_{n})=\sum_{i} C_{in}\chi_{i}(x)$.
One advantage of this choice of basis is that for small propagation
times, $C_{in}$ will have comparable amplitudes for all n, making the
associated linear system easier to solve with high accuracy.

Having chosen a temporal basis, coefficients $C_{in}$ which satisfy
Equation \ref{eq:stationaryphase} as well as the initial condition
$C_{i0}=\braket{\chi_{i}|\psi(t_{0})}$ can be found using Lagrange
multipliers.  If $S$ is the action accumulated in the time interval,
let $S'=S+\sum_{i}\lambda_{i} f^{*}_{i}$, where
$f_{i}=C_{i0}-\braket{\chi_{i}|\psi(t_{0})}$.  The least action
coefficients are found by minimizing $S'$ with the constraint that
$f_{i}=0$ for all $i$.  This yields a system of linear equations 
\begin{equation}
\sum_{i,n} C_{in}[i
  O_{ij}Q_{nm}-H_{ij}U_{nm}-V_{ijnm}]+\lambda_{j}=0
\label{eq:lalinearsystem}
\end{equation}
for all $j,m$, and
\begin{equation}
C_{i0}=\braket{\chi_{i}|\psi(t_{0})}
\label{eq:lainitialconditions}
\end{equation}
for all $i$.  The Lagrange multipliers $\lambda_{j}$ calculated in
this procedure are not needed by the propagator and can be discarded
after solving the linear system.

%

\section{Comparison with the Lanczos Propagator}
The least action propagator derived in the previous section is the
unique, variationally optimum propagator for a particular order in
time.  As such, it represents a formal improvement over all propagators
approximating the wavefunction as a low order polynomial in time --
forward and backwards Euler, Crank Nicholson, second order
differencing, etc.  However, it is less clear how this formal
improvement translates to a practical benefit, or how the least action
propagator compares to
methods which attempt to diagonalize the Hamiltonian in a Krylov
subspace, such as the popular short iterative Lanczos method \cite{lanczos}.

The Lanczos method works by repeatedly multiplying the initial
wavefunction by the Hamiltonian matrix to create a Krylov space of
limited dimension in which the matrix exponential $e^{-iHt}$ can be
calculated exactly.  It is considered to be both efficient and quickly
converging\cite{lanczoscomparison}.  Existing variational propagators
have focused on 
the evolution of the wavefunction in the Krylov subspace, yielding
convergence properties similar to the Lanczos
propagator\cite{triozon,talezer}.  The Chebyshev propagator, which
also uses repeated multiplication by the Hamiltonian matrix to
construct a Krylov space, converges similarly to the Lanczos
method\cite{chen}.

In the limit that the Krylov space has dimension equal to the full
Hamiltonian, the Lanczos method is equivalent to diagonalizing the
Hamiltonian, yielding $L=i\frac{d}{dt}\psi-H\psi=0$ at all times.
This solution is the global minimum action solution, and cannot be
improved upon.  However, for most applications, the Krylov subspace is
chosen to have a much smaller order -- typically in the range 1-10.

Because the Krylov subspace is constructed through repeatedly
multiplying an initial wavefunction by the Hamiltonian matrix, later
Krylov basis vectors will tend to some overlap with those eigenvectors
of $H$ with the largest eigenvalues.  For problems with Coulomb
singularities or fine spatial bases, it is not uncommon for these
largest eigenvalues to be artifacts of the choice of basis,
having no counterpart in the system being described.  However, they
may nonetheless serve to limit the stepsize which may be taken with
high accuracy.

If an ideal wavefunction (ie, without reference to a basis)
$\psi(x,t)$ can be expanded in terms of energy eigenfunctions over a
short time interval
\begin{equation}
\psi(x,t)=\sum_{\alpha} c_{\alpha}
f_{\alpha}(x)
e^{-i E_{\alpha}(t-t_{0})}
\end{equation}
and the evolution of the wavefunction's representation in the Krylov
subspace is given by
\begin{equation}
\phi(x,t)=\sum_{\beta} d_{\alpha,\beta}
g_{\beta}(x)
e^{-i E_{\beta}(t-t_{0})},
\end{equation}
where
$d_{\alpha,\beta}=c_{\alpha}\braket{g_{\beta}|f_{\alpha}}$,
then the error is given by
\begin{equation}
\delta(x,t)=\sum_{\alpha,\beta}d_{\alpha,\beta}g_{\beta}(x)(e^{-i
  E_{\alpha}(t-t_{0})}-e^{-i E_{\beta}(t-t_{0})})
\end{equation}
and
\begin{equation}
\braket{\delta|\delta}=\sum_{\alpha,\beta}|d_{\alpha,\beta}|^{2}
4\sin^{2}(\frac{E_{\alpha}-E_{\beta}}{2}(t-t_{0})).
\end{equation}

If the Krylov subspace is now partitioned into a ``good'' subspace
with $E_{\beta}<E_{\text{cutoff}}$ and a ``bad'' subspace with
$E_{\beta}>E_{\text{cutoff}}$, the error can be estimated by setting
$E_{\alpha}-E_{\beta}=0$ in the good subspace and
$E_{\beta}-E_{\alpha}=E_{H}$, where $E_{H}$ reflects the largest
eigenvalues of H, yielding
\begin{equation}
\braket{\delta|\delta}\approx\sum_{\alpha,\beta,
  E_{\beta}>E_{\text{cutoff}}}|d_{\alpha,\beta}|^{2} 
4\sin^{2}(\frac{1}{2}E_{H}(t-t_{0})).
\end{equation}
The error of the Lanczos method is thus minimized either when the
projection into the bad subspace is small or when $E_{H} \Delta t<<1$.

In contrast to the Lanczos method, the error of the least action
propagator is bounded by the error of the Taylor series of the true
wavefunction.  Thus
\begin{equation}
\braket{\delta|\delta}\le \sum_{\alpha} |c_{\alpha}|^{2}|e^{-i
  E_{\alpha}(t-t_{0})}-\sum_{n=0}^{N_{\text{max}}}\frac{(-iE_{\alpha})^{n}}{n!}(t-t_{0})^{n}|^{2}
\end{equation}
and the condition for the error to remain small is simply that
$E_{\text{cutoff}} \Delta t <1$.  For a problem with
$E_{\text{cutoff}}<<E_{H}$, the least action propagator offers the
possibility of much larger stepsizes at high accuracy than the Lanczos
propagator.

The two propagators were tested numerically using a 1 dimensional
Coulomb potential $1/x$ for $x$ ranging from 0 to 10.  The region was
separated into 100 finite element regions, with two quadratic finite
elements per region.  The wavefunction was restricted to have zero
value at both endpoints.  The largest eigenvalue of the resulting
Hamiltonian matrix was 499 Hartree.  The initial wavefunction was
chosen to be a Gaussian of unit width, centered at x=2.  The choice of
initial wavefunction and potential were made to ensure that the
wavefunction would be far from equilibrium and have a strong
interaction with the Coulomb potential, as for an electron wavepacket
scattering from a positive ion.

The accuracy of the Lanczos- and least action propagator was
calculated by propagating the initial wavefunction a single timestep
and comparing the resulting wavefunction with the ``true''
wavefunction found by directly diagonalizing the full
Hamiltonian. yielding an error
\begin{equation}
|err|=\braket{\delta(t+\delta t)|\delta(t+\delta t)}.
\end{equation}
For the least action propagator, the order of propagation is one less than
the degree of the polynomial basis functions.  For
the Lanczos propagator, the order is given by the dimension of the
Krylov subspace, starting with 0 for the initial wavefunction.

\begin{figure}
\begin{center}
\leavevmode
\includegraphics*[width=0.45\textwidth]{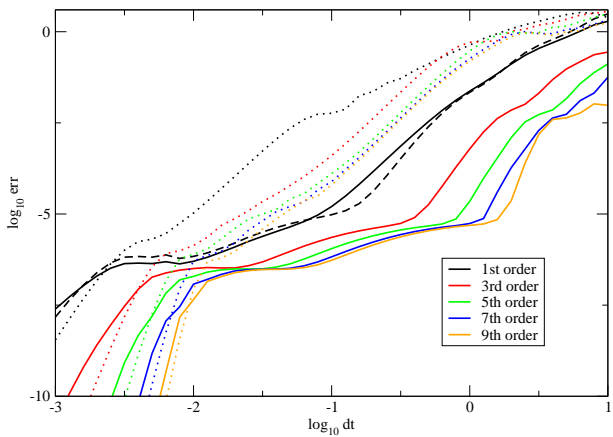}
\end{center}
\caption{(Color online) Point error $\braket{\delta(x,t+\Delta
    t)|\delta(x,t+\Delta t)}$ of Crank Nicholson (dashed line),
  Lanczos (dotted lines) and least action (solid 
  lines) propagators, as a function of step size.  For large
  stepsizes, the least action propagator is many times more accurate
  than the Lanczos propagator of the same order.}

\label{fig:errvstimestep}
\end{figure}

The error as a function of order and stepsize for the two methods is
shown in Figure \ref{fig:errvstimestep}.  Also shown in the figure is
the error vs time for the popular Crank Nicholson first order
propagator \cite{cranknicholson}.  As opposed to the global
error which was used in the derivation of the least action propagator,
these figures show the point error after a single propagation step.

These results show that the least action propagator offers the
potential for large timesteps to be taken with high accuracy, with the
greatest advantage coming from propagation at high order.  At first
order, the least action propagator gives approximately the same point
error as the Crank Nicholson method, while higher orders rapidly
decrease the error for a particular timestep, or alternatively
increase the size of the timestep which can be taken for a particular
desired error.  As the order increases, the error begins to saturate
as different order propagators converge on the same result.  That this
saturation does not result in zero error may result from numerical
error in the diagonalization of the Hamiltonian or the linear solver.

For all orders tested, the least action propagator was many times more
accurate than the Lanczos 
propagator for large stepsizes.  For very small stepsizes, the error
of both methods was comparable, with the Lanczos method more
accurate.  Both methods became much more accurate for stepsizes of
less than $~10^{-2}$, which is interpreted to mean that the initial
wavefunction had some projection onto very high energy
eigenstates; ie, the sample problem did not have
$E_{\text{cutoff}}<<E_{H}$.

\begin{figure}
\begin{center}
\leavevmode
\includegraphics*[width=0.45\textwidth]{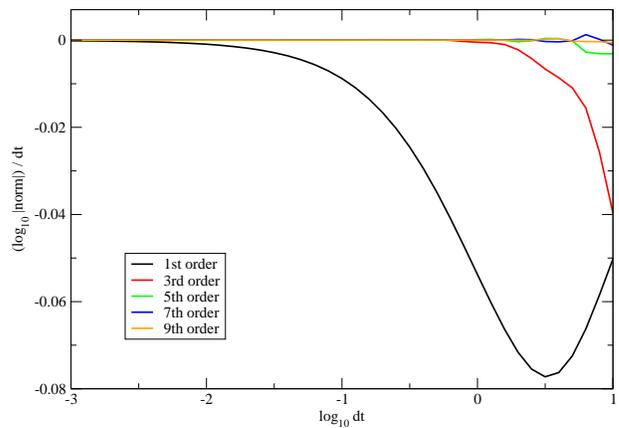}
\end{center}
\caption{(Color online) Exponential growth rate
  $(\log_{10}|\text{norm}|)/dt$ vs propagation time for different
  orders of the least action propagator.  Higher orders show a growth
  rate closer to zero.}

\label{fig:normerrvstimestep}
\end{figure}

One weakness of the least action propagators which arises from the
choice of polynomial basis functions is that the norm of the
propagated wavefunction is not required to be a constant as a function
of time.  Figure \ref{fig:normerrvstimestep} shows the rate of
growth/decay of the norm $\braket{\phi|\phi}$ for different orders
of propagation as a function of step 
size.  Here the largest deviation from zero growth rate is found for
the combination of low order and large stepsize.  Higher order
propagators show growth rates close to zero.   For problems
requiring repeated use 
of the propagator over many timesteps, it is thus likely that the
propagated wavefunction will need to be renormalized periodically.
Because the norm is still very close to 1, such renormalization has a
minimal effect on the point error shown in Figure
\ref{fig:errvstimestep}.

From these figures, it is apparent that the least action propagator
works best at high orders, which offer the combination of large time
steps, high accuracy and low rates of growth or decay of the norm.
However, high order also increases the size of the linear system which
must be solved at each step.  For a basis consisting of $n_{x}$
spatial and $n_{t}$ temporal basis functions, the least action
propagator requires $n_{s}=n_{x}(n_{t}+1)$ variables to be solved
for.  While specific implementations are beyond the scope of this
paper, the question of how best to solve this linear system will play
a crucial role in applying the least action propagator to real world
problems.  To this end, a few features of the least action linear
system are worth noting.  First among these is the highly separable
nature of the least action linear system defined in equations
\ref{eq:lalinearsystem} and \ref{eq:lainitialconditions}.  In equation
\ref{eq:lalinearsystem}, the variation of the action is given as the
sum of three matrices, $O_{ij}Q_{nm}$, $H_{ij}U_{nm}$, and
$V_{ijnm}$.  Of these, the first two are separated into the product
of spatial and temporal matrices.  Because of this, these two matrices
inherit the sparsity and/or banded structure of the underlying spatial
matrices.  The case is similar for the nonseparable $V_{ijnm}$: the
integral over time and space is nonzero only if the integral over
space is nonzero.  Because of this, basis sets 
such as finite elements which are chosen for the structure of their
Hamiltonian and overlap matrices will retain these advantages in the
least action equation.  For a banded problem such as the 1D finite
element problem treated in this paper, the bandwidth of the linear
system increases linearly with the order of the propagator, giving an
overall $n_{t}^{2}$ scaling with the order.

For very large problems which lack such a simple structure, it is
likely that solution of the least action linear system will require
use of an iterative solver, such as those available in the PETSc
\cite{petsc} or Trilinos \cite{trilinos} libraries.  Such solvers,
require calculating a matrix vector product at every 
iteration.  Here the advantages of the least action equation's
separable form are very 
apparent, particularly in the case of a static Hamiltonian.  A single
matrix vector product of the linear system defined in equations
\ref{eq:lalinearsystem} and \ref{eq:lainitialconditions} requires 1
(very expensive) matrix-vector multiplication by the unseparated
matrix $V_{ijnm}$, $2n_{t}$  independent (expensive) multiplications
of the form $D_{im}=M_{ij}C_{jm}$,
 where $M=O$ or $H$, followed by
$2n_{x}$ (cheap) independent multiplications of the form $F_{in}=N_{nm}D_{im}$,
where $N=Q$ or $U$.  Thus, although the linear system which must be
solved is very large, it is well suited to iterative solution.  The
overall scaling of one iteration with respect to propagator order will
be limited by the slowest of these steps, which may depend on
specifics of the data structure and architecture of the system used.

This paper has addressed the problem of propagating a wavefunction in
time by minimizing he accumulated action.  For a particular choice of
spatial and temporal basis functions, the problem is reduced to
solution of a (potentially very large) system of linear equations.
This linear system inherits the sparsity and/or banded structure of
the spatial Hamiltonian and overlap matrices, while its separable
structure makes it amenable to solution by iterative solvers.  The
resulting propagator was shown to have improved convergence relative
to the commonly used short iterative Lanczos propagator, giving the
potential for larger stepsizes at high accuracy.  

The derivation of the action as a measure of propagation error is a
powerful result which offers many opportunities for the systematic
improvement of propagation schemes.  By monitoring the spacetime
volumes where the most action is accumulated, spatial or temporal
bases could be selectively refined to increase the total accuracy of a
propagation step at minimal additional computational effort.  For this
reason, the full power of the action minimization principle may be yet
to be unlocked.

\bibliographystyle{phcpc}

\end{document}